\documentclass[12pt]{article}
\usepackage{amsfonts,amsmath,amssymb}
\usepackage{color}
\usepackage{epic}
\usepackage{graphics}
\usepackage{boxedminipage}
\usepackage{epsfig}
\usepackage{changebar}
\usepackage{lscape}

\numberwithin{equation}{section}

\newcommand{\invdots}{{\mathinner{\mkern1mu\raise1pt
   \vbox{\kern7pt\hbox{.}}\mkern2mu\raise4pt\hbox{.}
   \mkern2mu\raise7pt\hbox{.}\mkern1mu}}}

\newtheorem{definition}{Definition}[section]

\newtheorem{proposition}[definition]{Proposition}

\newcommand{\prf}{\underline{Proof:}\ }
\newcommand{\finprf}{\null \hfill {\rule{5pt}{5pt}}\\ \null}

\definecolor{dcyan}{rgb}{0,.8,.8}
\definecolor{ddcyan}{rgb}{0,.6,.6}
\definecolor{dgreen}{rgb}{0,.8,0}
\definecolor{lyellow}{cmyk}{0,0,.2,0}
\definecolor{rose}{rgb}{1,.6,.6}
\definecolor{violet}{rgb}{.9,0,.6}

\newcommand{\1}{\mbox{\hspace{.0em}1\hspace{-.2em}I}}

\def\CC{{\mathbb C}}

\def\RR{{\mathbb R}}

          \def\cB{{\cal B}}

                    \def\cL{{\cal L}}

          \def\cT{{\cal T}}

\newcommand{\lda}{\lambda}

\newcommand{\ie}{{\it i.e.}\ }

\def\tU{{\widetilde{U}}}
\def\tV{{\widetilde{V}}}
\def\tW{{\widetilde{W}}}
\def\tq{{\tilde{q}}}
\def\tr{{\tilde{r}}}
\def\tu{{\tilde{u}}}
\def\tv{{\tilde{v}}}

\newcommand{\be}{\begin{eqnarray}}
\newcommand{\ee}{\end{eqnarray}}

\begin{document}

\newpage
\pagestyle{empty} \setcounter{page}{0}
%%%%%%%%%%%%%%%%%%%%%%%%%%%%%%%%
%%%%%  HEADINGS POUR DRAFT  %%%%%%

\markright{\today\dotfill DRAFT SOLIM\dotfill }
%\pagestyle{myheadings}
%%%%%%%%%%%%%%%%%%%%%%%%%%%%%%%%

\vspace{20mm}

\begin{center}
   {\LARGE  {\sffamily On a systematic approach to defects in classical integrable field theories
  }}\\[1cm]

\vspace{10mm}

{\large \textbf{V. Caudrelier}\footnote{E-mail: v.caudrelier@city.ac.uk}}\\
\vspace{1cm}
City University\\
Centre for Mathematical Science\\
Nothampton Square\\
LONDON EC1V 0HB\\
UK
\end{center}
\vfill

\begin{abstract}
We present an inverse scattering approach to defects in classical integrable field theories. Integrability
is proved systematically by constructing the generating function of the infinite set of modified integrals
of motion. The contribution of the defect to all orders is explicitely identified in terms of a defect 
matrix. The underlying geometric picture is that those defects correspond to B\"acklund transformations localized at a 
given point. A classification of defect matrices as well as the corresponding defect conditions is performed. 
The method is applied
to a collection of well-known integrable models and previous results are recovered (and extended) 
directly as special cases. Finally, a brief discussion of the classical $r$-matrix approach in this context
shows the relation to inhomogeneous lattice models and the need to resort to lattice regularizations of integrable field 
theories with defects.
\end{abstract}

\vfill \centerline{PACS numbers: 02.30.Ik, 02.30.Jr, 02.30.Zz, 11.10.Kk, 11.10.Lm}

\vfill
%\rightline{\tt math-ph/yymmnnn}

\newpage
\pagestyle{plain}
%%%%%%%%%%%%%%%%%%%%%%%%%%%%%%%%
%%%%  FIN PAGE DE TITRE  %%%%%%
%%%%%%%%%%%%%%%%%%%%%%%%%%%%%%
\setcounter{footnote}{0}

\section*{Introduction}

The topic of defects, or impurities, in integrable systems has quite a rich literature, especially for quantum aspects
\cite{DMS,KL,S,CFG,MRS,CMR,BG,CMRS}, even if quite a lot remains to be done. Strangely enough, the problem of integrable
defects in classical field theories had received less attention. The pioneering paper \cite{BT1} is worth mentioning for
the introduction of a so-called "spin impurity" in 
the nonlinear Schr\"odinger equation as a first step to tackle the problem on the half-line
with integrable boundary conditions. This topic has been revived recently in the series of papers by P. Bowcock, E. Corrigan and 
C. Zambon \cite{BCZ,CZ}. The lagrangian formalism is used in all these papers to described integrable field theories with
internal boundary conditions interpreted as the presence of a defect. The defect conditions emerge from
a local lagrangian density concentrated at some fixed point and are obtained from a variational argument. The question 
that is addressed is then: how to select conditions which leave the full theory integrable? The common underlying philosophy 
is to impose that a modified momentum, taking into account the presence of the defect, should be a conserved quantity
while the breaking of translation invariance obviously entails that the bulk momentum will not be conserved. It turns out
that this idea allows to pick up certain classes of defect lagrangian. Then, a general argument for integrability 
is based on the construction of modified
Lax pair involving a limiting procedure. It is checked explicitely for a few conserved charges of certain models.
One must note the nice observation made in each case: this procedure yields frozen B\"acklund transformations
as the defect conditions for the fields.

The object of this paper is to unify the results obtained by this case by case approach. We take advantage 
of the common features that have been observed.
To do so, we use the efficient inverse scattering method formalism (instead of the lagrangian formalism) and 
implement defect conditions corresponding to frozen B\"acklund transformations. It is important to note that
the role of B\"acklund transformations as a means to generate integrable boundary-initial value systems solvable
by inverse scattering method has been discovered and used in \cite{H, BT2, T}. The idea was to fold two copies of the original integrable 
system related by B\"acklund transformations by using compatible reductions on the fields (for example $u(x)=u(-x)$).
Here, we do not fold and use the fact that B\"acklund transformations have a very nice formulation in the inverse 
scattering method. They can be encoded in matrices, representing gauge
transformations of the underlying auxiliary problem and, in the present context, giving rise to defect matrices. 
Thanks to this formulation, we are able
to prove systematically the existence of an infinite set of modified conservation laws, ensuring integrability. 
The main result of this paper is the explicit identification of the 
generating function of the defect contributions at all orders, \ie for any conserved charge, for any 
integrable evolution equation of the AKNS \cite{AKNS} or KN \cite{KN} schemes of the inverse scattering method. 
This provides an efficient algorithm to compute the modified conserved quantities, given the defect matrices.
One of the 
advantage of the method is that the proof of integrability does not require any modification of the usual Lax pair
formulation for integrable field theories. Another is that there is no guess work for finding the defect contributions.
They are obtained from a classification of defect matrices.

The paper is organized as follows. In Section 1, the general auxiliary problem formalism we use is presented. 
We establish our main results about the infinite set of conservation laws in the presence of a 
defect. The generalization to several defects is also explained. In Section 2, the defect matrices are classified
within a certain class of gauge transformations of the auxiliary problem.
In section 3, we illustrate our systematic method on several well-known examples of integrable nonlinear equations. They
correspond to all the classical field theories that have been explored in the lagrangian formalism (with the 
exception of the affine Toda field theories). For these models, all the previous results
are recovered (and even generalized) and are extended to higher orders. Section 4 is devoted
to the extension of the method to another inverse scattering method scheme, the Kaup-Newell scheme \cite{KN}, which
describes other classes of integrable nonlinear equations including e.g. the derivative nonlinear Schr\"odinger
equation. In section 5, we discuss in more detail the question of integrability of such models with defects. It is argued that our approach
allows to make a connection between the lagrangian approach and the classical $r$-matrix formalism. This requires the use 
of lattice regularizations. Our conclusions and perspectives for future investigations are 
gathered in the last section.

\section{General settings and results}

\subsection{Lax pair formulation}
In the AKNS scheme \cite{AKNS}, an integrable evolution equation on the 
line can be formulated
as a compatibility condition, or zero curvature condition, of a linear differential problem for
an auxiliary wavefunction $\Psi(x,t,\lda)$ involving two $2\times 2$ matrix-valued functions 
$U(x,t,\lda)$ and $V(x,t,\lda)$ such that
\be
\label{pb_aux}
\begin{cases}
\Psi_x=U\Psi\,,\\
\Psi_t=V\Psi\,,
\end{cases}
\ee
where the subcripts $x$ and $t$ denote differentiation with respect to these variables.
In the rest of the paper, we will drop the arguments whenever this is not misleading.
The parameter $\lda$ is called spectral parameter. Then, for appropriate choices of $U$ and $V$, 
the integrable system at hand is equivalent to the
compatibility condition $\Psi_{xt}=\Psi_{tx}$ giving rise to the so-called \textit{zero curvature 
condition}
\begin{eqnarray}
\label{zero_curv}
\forall\,\lda~~,~~U_t-V_x+[U,V]=0\,.
\end{eqnarray}
Large classes of integrable nonlinear evolution equations can be described this way
among which some of the most famous are the cubic nonlinear Schr\"odinger (NLS), sine/sinh-Gordon (sG), Liouville, 
Korteweg-de Vries (KdV) or its modified version (mKdV).
It is known that this presentation allows one to construct generically the infinite set of conservation 
laws associated to the integrable equation. For self-containedness, we recall the main ideas. Let us fix $U$ and $V$ to be 
$2\times 2$ traceless matrices as follows
\begin{eqnarray}
\label{lax_pair}
U=\left(\begin{array}{cc}
-i\lda & q\\
r & i\lda
\end{array}\right)\equiv -i\lda\sigma_3+
W~~,~~V=\left(\begin{array}{cc}
A & B\\
C & -A
\end{array}\right)\,,
\end{eqnarray}
where $q(x,t)$ and $r(x,t)$ are the fields satisfying the evolution equation. In this paper, we will
fix the class of solutions to be that of sufficiently smooth \cite{NB} decaying fields\footnote{The decaying properties are chosen
so as to ensure certain analytic properties of the scattering data for (\ref{pb_aux}), see e.g. \cite{AKNS}.
Typically, a polynomial decay is sufficient.} as $|x|\to\infty$ and 
the following behaviour is assumed\footnote{The only
exception is the Liouville equation for which no specific boundary condition is assumed.}
\be
\label{BC}
A(x,t,\lda)\to \omega(\lda)~~,~~B(x,t,\lda),\,C(x,t,\lda)\to 0~\text{as}~|x|\to\infty\,.
\ee
The vector-valued function $\Psi$ is split as 
\begin{eqnarray}
\Psi=\left(\begin{array}{c}
\Psi_1 \\
\Psi_2
\end{array}\right)\,.
\end{eqnarray}
Let $\Gamma=\Psi_2\Psi_1^{-1}$, then the identification of the infinite set of conservation laws 
follows from a conservation equation
\begin{eqnarray}
\label{conservation}
\left(q\Gamma\right)_t&=&\left(B\Gamma+A\right)_x\,,
\ee
and a Ricatti equation for $\Gamma$ 
\be
\label{Ricatti_x}
\Gamma_x=2i\lda\Gamma+ r-q\Gamma^2\,.
\ee
This equation follows directly from the $x$ part of (\ref{pb_aux}). The conservation equation is obtained
from the $t$ part to get $\Gamma_t$, from the $(12)$ element of (\ref{zero_curv}) to get $q_t$ and combining
with (\ref{Ricatti_x}) and with the $(11)$ element of (\ref{zero_curv}) giving $A_x$.

Thus, expanding $\Gamma$ as $\lda\to\infty$
\be
\Gamma=\sum_{n=1}^\infty\frac{\Gamma_n}{(2i\lda)^n}\,,
\ee
the conserved quantities read
\be
I_n=\int_{-\infty}^\infty q\Gamma_n dx~~,~~n\ge 1\,,
\ee
where
\be
\Gamma_1=-r~~,~~\Gamma_{n+1}=\Gamma_{nx}+q\sum_{k=1}^{n-1}\Gamma_k\Gamma_{n-k}~~,~~n\ge 1\,.
\ee

\subsection{Implementing defect boundary conditions}

Generally speaking, a defect in $(1+1)$-dimensional integrable field theories can be viewed
as internal boundary conditions on the field and its time and space derivatives at a given point on the line. 
In other words, one wants to glu together two 
solutions of the evolution equation in a specific way and at a particular point. To this end, 
let us consider another copy of the auxiliary problem. We introduce another Lax pair 
$\widetilde{U}$, $\widetilde{V}$ defined as in (\ref{lax_pair}) and (\ref{BC}) with $q$, $r$
replaced by $\tq$, $\tr$. We consider the analog of (\ref{pb_aux}) for
\begin{eqnarray}
\label{def_L}
\widetilde{\Psi}(x,t,\lda)=L(x,t,\lda)\Psi(x,t,\lda)\,.
\end{eqnarray}
The matrix valued function $L(x,t,\lda)$ satisfies the following partial differential equations 
for any $x$ and $t$,
\begin{eqnarray}
\label{eq_diff_L}
L_x&=&\widetilde{U}L-LU\,,\\
\label{eq_diff_L2}
L_t&=&\widetilde{V}L-LV\,.
\end{eqnarray}
In this paper, we want to think of this matrix as generating
the defect conditions at a specific point, $x_0$ say. Following the terminology of \cite{BCZ,CZ}, $L$ is called
the \textit{defect matrix}. We present a classification of the simplest nontrivial such matrices in the next section.

Now we turn to the general contruction of
the generating function of the infinite set of modified conservation laws due to the presence of
a defect.
This is the main
result of this paper and establishes integrability for any nonlinear integrable equation
of the AKNS scheme with a defect realizing a frozen B\"acklund transformation. 
For particular models (NLS, sG, Liouville, KdV and mKdV), 
it proves to all orders the results of \cite{BCZ,CZ} 
about the defect contribution
and gives an explicit form for it. In addition, this is done without resorting to
a modified Lax pair formalism involving a complicated limiting procedure to construct the conserved charges.
To illustrate this, we will discuss those particular examples in Section \ref{examples}.

To fix ideas, 
we choose a point $x_0\in\RR$ and we suppose that the auxiliary problem (\ref{pb_aux}) exists
for $x>x_0$ while the one for $\tU$ and $\tV$ exists for $x<x_0$. We also assume that the two systems
are connected by the relations (\ref{eq_diff_L}) and (\ref{eq_diff_L2}) at $x=x_0$.
Then, the following holds
\begin{proposition}\label{prop_conserved}
The generating function for the integral of motions reads
\begin{eqnarray}
\label{generating}
I(\lda)&=&I_{bulk}^{left}(\lda)+I_{bulk}^{right}(\lda)+I_{defect}(\lda)\,,
\ee
where\be
I_{bulk}^{left}(\lda)&=&\int^{x_0}_{-\infty}
\tq\widetilde{\Gamma}dx\,,\\
I_{bulk}^{right}(\lda)&=&\int_{x_0}^\infty q\Gamma dx\,,\\
I_{defect}(\lda)&=&-\ln (L_{11}+L_{12}\Gamma)|_{x=x_0}\,,
\end{eqnarray}
and $L_{ij}$'s are the entries of the defect matrix $L$.
\end{proposition}
\prf
From the general result (\ref{conservation}), we get
\begin{eqnarray}
\label{conservation1}
\left(q\Gamma\right)_t&=&\left(B\Gamma+A\right)_x~~,~~\forall x>x_0\,,\\
\label{conservation2}
\left(\tq\widetilde{\Gamma}\right)_t&=&\left(\widetilde{B}
\widetilde{\Gamma}+\widetilde{A}\right)_x~~,~~\forall x<x_0\,,
\end{eqnarray}
where $\widetilde{\Gamma}$ is defined from $\widetilde{\Psi}$ as in the previous section, $\widetilde{\Gamma}
=\widetilde{\Psi}_2\widetilde{\Psi}_1^{-1}$.
From this and the rapid decay of the fields, we get
\begin{eqnarray}
\label{intermed}
\partial_t\int_{x_0}^\infty q\Gamma dx+\partial_t\int^{x_0}_{-\infty}
\tq\widetilde{\Gamma}dx=\left(\widetilde{B}
\widetilde{\Gamma}+\widetilde{A}-(B\Gamma+A)\right)|_{x=x_0}
\end{eqnarray}
The crucial point now is that the right-hand-side is a total time derivative of a quantity
evaluated at $x=x_0$: it is the contribution of the defect to the conserved quantities as we now show.
From (\ref{def_L}) we get $\widetilde{\Gamma}=(L_{21}+L_{22}\Gamma)(L_{11}+L_{12}\Gamma)^{-1}$.
Then, using (\ref{eq_diff_L2}) at $x=x_0$ to eliminate $\widetilde{A}$ and $\widetilde{B}$, one gets
\begin{eqnarray}
&&\hspace{-2cm}\left(\widetilde{B}
\widetilde{\Gamma}+\widetilde{A}-(B\Gamma+A)\right)|_{x=x_0}\\
&=&
\left\{\partial_tL_{11}+\partial_tL_{12}\Gamma+L_{12}(C-2A\Gamma-B\Gamma^2)\right\}
(L_{11}+L_{12}\Gamma)^{-1}|_{x=x_0}\,.
\end{eqnarray}
The final step consists in noting that the $t$ part of (\ref{pb_aux}) implies another
Ricatti equation
\be
\label{ricatti_t}
\Gamma_t=C-2A\Gamma-B\Gamma^2\,,
\ee
so that
\begin{eqnarray}
\partial_t\int_{x_0}^\infty q\Gamma dx+\partial_t\int^{x_0}_{-\infty}
\tq\widetilde{\Gamma}dx=\frac{(L_{11}+L_{12}\Gamma)_t}{
(L_{11}+L_{12}\Gamma)}|_{x=x_0}\,,
\end{eqnarray}
Therefore, 
\be
\partial_t I(\lda)=0\,.
\ee
\finprf
Note that this holds in all generality for any integrable evolution equation in the 
AKNS scheme with decaying fields. The latter can be relaxed for some models, and
we discuss in detail below the Liouville equation for which this does not hold.

\subsection{Several defects}\label{several}

It is quite straightforward to repeat the general argument for the construction of 
conservations laws in the case of several defects. Suppose we have $N+1$ auxiliary problems
connected two by two at points $x_1,\dots,x_N$ by matrices $L^1,\dots,L^N$. Then, it is easy
to see that the contribution of these $N$ defects is the sum of the contributions from each defect. 
Indeed, using $x_0=-\infty$, $x_{N+1}=+\infty$ and otherwise obvious notations, 
the generating function for the integral of motion reads
\begin{eqnarray}
I(\lda)&=&\sum_{j=1}^{N+1} I_{bulk}^{j}(\lda)+\sum_{j=1}^{N}I_{defect}^{j}(\lda)\,,\\
I_{bulk}^{j}(\lda)&=&\int_{x_{j-1}}^{x_j}
q_j\Gamma_j dx~~,~~j=1,\dots,N+1\,,\\
I_{defect}^{j}(\lda)&=&\ln ((L_{j})_{11}+(L_{j})_{12}\Gamma_{j})|_{x=x_0}~~,~~j=1,\dots,N\,.
\end{eqnarray}

\section{Defect matrices}

In this section, we derive a large class of defect matrices satisfying (\ref{eq_diff_L}) and (\ref{eq_diff_L2}) together
with the associated conditions they entail on the fields: the B\"acklund transformations. The latter will
become the defect conditions when imposed at $x=x_0$.

\subsection{Generalities}

The matrix $L$ preserves the zero curvature condition as is easily seen by writing $L_{xt}=L_{tx}$,
\be
(\forall\,\lda\,,\,U_t-V_x+[U,V]=0)\Leftrightarrow (\forall\,\lda\,,\,\tU_t-\tV_x+[\tU,\tV]=0)\,.
\ee
In other words, if $q$, $r$ are solutions of the evolution equation described by $(U,V)$ then $\tq$, $\tr$
are solutions of the evolution equation described by $(\tU,\tV)$ under the transformation induced by $L$ and vice versa. 
This is just the usual definition of a B\"acklund transformation and this shows the connection with
the idea of frozen B\"acklund transformations discussed above. In other words, we look for defect matrices 
in the class of matrices realizing B\"acklund transformations between two nonlinear integrable evolution equations. 
Note that the evolution equations need not be the same in
general. If they are, the terminology auto-B\"acklund transformation is usually used. Such matrices
are sometimes referred to as Darboux matrices (see e.g. \cite{RS}). Even if a lot is known on these matrices, we proceed
with their derivation in the form needed for this paper. We adopt a pedestrian method which does not require any previous 
knowledge of their theory. In particular, no reference to the wavefunction of the auxiliary problem or to a 
special Riemann problem is needed (which are usually the methods encountered in the literature).

Let us establish some general facts about $L$. First, there is some freedom in its normalization
 coming from the invariance of the zero-curvature condition under the transformation $(U,V)\to
(M^{-1}UM, M^{-1}VM)$ for any invertible matrix $M$ independent of $x$ and $t$. This also obviously preserves 
the tracelessness property. In particular, left multiplication
of $L$ by $M^{-1}$ amounts to apply this transformation to $(\tU,\tV)$ while right multiplication by $M$ applies it
to $(U,V)$. Then, we have the
\begin{proposition}\label{prop2}
The determinant of $L$ is independent of $x$ and $t$
\be
\det L(x,t,\lda)=f(\lda)\,.
\ee
\end{proposition}
\prf
The result follows from the Jacobi formula
\be
\left(\det L(x,t,\lda)\right)_x&=&\det L(x,t,\lda)\text{Tr}(\tU-U)\,,\\
\left(\det L(x,t,\lda)\right)_t&=&\det L(x,t,\lda)\text{Tr}(\tV-V)\,,
\ee
and the tracelessness of $U,\tU,V,\tV$.
\finprf

At this stage, it is hard to go further without specifying $U,\tU,V,\tV$ a bit more. 
Let us simply note that given $U,\tU,V,\tV$ and initial-boundary values for the fields, the integration of 
(\ref{eq_diff_L}) and (\ref{eq_diff_L2}) gives for instance (the path of integration being irrelevant)
\be
L(x,t,\lda)=L(x_0,t_0,\lda)+\int_{x_0}^{x}(\tU L-LU)|_{\tau=t} dy +
\int_{t_0}^{t}(\tV L-LV)|_{y=x_0} d\tau\,.
\ee
The formal iteration of the previous equation suggests
 that, in general, $L$ has a complicated Laurent series structure as a function of $\lda$. 
In the following, we will assume that $L$ has only a finite number of terms and,
recalling that it is defined up to a scalar function in $\lda$, we will look for a solution of the form
\be
\label{truncate}
L(x,t,\lda)=\sum_{n=0}^NL^{(n)}(x,t)\lda^{-n}\,.
\ee
Actually, we shall consider the case $N=1$ which we study in detail. 
For convenience, we also restrict our attention to auto-B\"acklund matrices. We comment later on on the fact that this is not
necessary in our approach, one of the crucial ingredient being simply 
that the evolution equations have the same dispersion 
relation (see (\ref{limit_VV}) below).

\subsection{Construction for N=1}
The defect matrix is of the form\footnote{Note that it is implicitly
assumed that both $L^{(0)}$ and $L^{(1)}$ are not trivial since otherwise, the B\"acklund
transformation is essentially the trivial one $\tW=W$.}
\be
\label{forme_L}
L(x,t,\lda)=L^{(0)}(x,t)+L^{(1)}(x,t)\lda^{-1}\,.
\ee
The defining relation (\ref{eq_diff_L}) is equivalent to 
\be
\label{x1}
0&=&\left[L^{(0)},\sigma_3\right]\,,\\
\label{x2}
L^{(0)}_x&=&i\left[L^{(1)},\sigma_3\right]+\tW L^{(0)}-L^{(0)}W\,,\\
\label{x3}
L^{(1)}_x&=&\tW L^{(1)}-L^{(1)}W\,.
\ee
If $V$,$\tV$ are polynomials in $\lda$ with coefficients $V^{(j)}(x,t)$, $\tV^{(j)}(x,t)$, 
$j=0,\cdots,N$, equation (\ref{eq_diff_L2}) is equivalent to 
\be
\label{t1}
L^{(1)}_t&=&\tV^{(0)}L^{(1)}-L^{(1)}V^{(0)}\,,\\
\label{t2}
L^{(0)}_t&=&\tV^{(1)}L^{(1)}-L^{(1)}V^{(1)}+\tV^{(0)}L^{(0)}-L^{(0)}V^{(0)}\,,\\
0&=&\tV^{(2)}L^{(1)}-L^{(1)}V^{(2)}+\tV^{(1)}L^{(0)}-L^{(0)}V^{(1)}\,,\\
&\vdots&\\
\label{tN}
0&=&\tV^{(N)}L^{(0)}-L^{(0)}V^{(N)}\,.
\ee
If $V$,$\tV$ are polynomials in $\lda^{-1}$, the equations are the same under the exchange 
$L^{(0)}\leftrightarrow L^{(1)}$. 

Let us make a few remarks. First, when we have found the matrix
 $L$, equations (\ref{x3}) and (\ref{t1}) will give the $x$ and $t$ parts of the 
 corresponding B\"acklund transformations for the fields. Then, in traditional 
 approaches, equation (\ref{x2}) is used to construct new soliton solutions, $\tW$, from 
 given solutions $W$ and the knowledge of the B\"acklund transformation. We will not discuss 
 this last step in this paper and refer the reader to the vast literature on the subject (see e.g.
 \cite{DJ} and references therein).
 
We now proceed with the statement of the general results of this
section.
\begin{proposition}\label{th} {\bf Defect matrix}\\
The defect matrix $L$ has the following general form
\be
L=\1_2+\lda^{-1}\left(\begin{array}{cc}
\frac{1}{2}\left\{\alpha_+ \pm\left[\alpha_-^2-4a_2a_3\right]^{1/2} \right\} & a_2\\
a_3 & \frac{1}{2}\left\{\alpha_+\mp\left[\alpha_-^2-4a_2a_3\right]^{1/2} \right\}
\end{array}\right)\,,
\ee
where
\be
a_2=-\frac{i}{2}(\tq-q)~~,~~a_3=\frac{i}{2}(\tr-r)\,,
\ee
and $\alpha_\pm\in\CC$ are the ($x,t$-independent) parameters of the defect.
\end{proposition}
\prf
Equation (\ref{x1}) implies that $L^{(0)}$ in (\ref{forme_L}) is diagonal and then, equation (\ref{x2}) shows that
the diagonal elements do not depend on $x$. Therefore, we can consider equation (\ref{t2}) as 
$|x|\to\infty$. Recall that we have
\be
\label{limit_VV}
V,\tV\to \omega(\lda)\,\sigma_3~~,~~|x|\to \infty\,.
\ee
Writing $\displaystyle\omega(\lda)=\sum_{n=0}^N\omega^{(n)}\lda^n$ and denoting $\displaystyle L_\infty^{(1)}=\lim_{|x|
\to\infty}L^{(1)}(x,t)$, we get
\be
L^{(0)}_t&=&\omega^{(1)}\left[\sigma_3,L_\infty^{(1)}\right]+\omega^{(0)}\left[\sigma_3,L^{(0)}\right]\\
&=&-i\omega^{(1)}\lim_{|x|\to\infty}(\tW L^{(0)}-L^{(0)}W)\\
&=&0\,,
\ee
where we have used equation (\ref{x1}) and (\ref{x2}) in the second equality and the fact that we consider decaying fields in 
the last equality. The proof is similar if $V$, $\tV$ are polynomials in $\lda^{-1}$. So, as explained above, we can 
left multiply with $\left(L^{(0)}\right)^{-1}$ and work with $L^{'}=\left(L^{(0)}\right)^{-1}L$. We drop the $'$ in the following
but remember that $L^{(0)}$ should now be $\1_2$ in all the equations (\ref{x1}-\ref{tN}). Next, denote
\be
L^{(1)}=\left(\begin{array}{cc}
a_1 & a_2\\
a_3 & a_4
\end{array}\right)\,,
\ee
and $\alpha_1$, $\alpha_2$ its eigenvalues.
Then, equation (\ref{x2}) gives immediately $a_2=-\frac{i}{2}(\tq-q)$, $a_3=\frac{i}{2}(\tr-r)$. Now,
the elements $a_1$ and $a_4$ are easily computed from 
\be
a_1a_4-a_2a_3=\alpha_1\alpha_2~~\text{and}~~a_1+a_4=\alpha_1+\alpha_2\,,
\ee
and introducing $\alpha_\pm=\alpha_1\pm\alpha_2$.
Finally, we need to show that $\alpha_\pm$ is independent of $x$ and $t$. Let $\ell_1$ and $\ell_2$ be the eigenvalues of $L$. 
It is enough to prove that $\ell_1$ and $\ell_2$ are independent of $x$ and $t$. In turn, it is sufficient to 
prove that $\ell_1\ell_2$ and $\ell_1+\ell_2$ are independent of $x$ and $t$. From proposition \ref{prop2}, we already 
know that $\ell_1\ell_2=f(\lda)$. Next, we prove $\ell_1+\ell_2=g(\lda)$. Suppose $\ell_1+\ell_2=g(x,t,\lda)$, then,
using (\ref{eq_diff_L}) and (\ref{x2})
\be
g_x&=&\text{Tr}\,\left[L(\tW-W)\right]\\
&=&\lda^{-1}\text{Tr}\,\left[L^{(1)}(\tW-W)\right]\\
&=&\frac{i}{\lda}\text{Tr}\,\left[L^{(1)}\left[\sigma_3,L^{(1)}\right]\right]\\
&=&0\,.
\ee
Now, $g_t=\text{Tr}\,\left[L(\tV-V)\right]$ can be evaluated as $x\to\infty$ for which we
know that $\tV-V\to 0$. So $g_t=0$.
\finprf

The B\"acklund transformations associated with the matrix $L$ read:
\begin{itemize}
\item For the $x$ part,
\be
\label{general_BT_x1}
a_{1x}&=&\tq a_3-ra_2\,,\\
\label{general_BT_x2}
a_{2x}&=&\tq a_4-qa_1\,,\\
\label{general_BT_x3}
a_{3x}&=&\tr a_1-ra_4\,,\\
\label{general_BT_x4}
a_{4x}&=&\tr a_2-qa_3\,,
\ee

\item For the $t$ part if $V$, $\tV$ are polynomials in $\lda$,
\be
\label{general_BT_t1}
a_{1t}&=&(\widetilde{A}^{(0)}-A^{(0)})a_1+\widetilde{B}^{(0)}a_3-C^{(0)}a_2\,,\\
\label{general_BT_t2}
a_{2t}&=&(\widetilde{A}^{(0)}+A^{(0)})a_2+\widetilde{B}^{(0)}a_4-B^{(0)}a_1\,,\\
\label{general_BT_t3}
a_{3t}&=&-(\widetilde{A}^{(0)}+A^{(0)})a_3+\widetilde{C}^{(0)}a_1-C^{(0)}a_4\,,\\
\label{general_BT_t4}
a_{4t}&=&-(\widetilde{A}^{(0)}-A^{(0)})a_4+\widetilde{C}^{(0)}a_2-B^{(0)}a_3\,,
\ee

\item For the $t$ part if $V$, $\tV$ are polynomials in $\lda^{-1}$,
\be
\label{general_BT_tt1}
a_{1t}&=&(\widetilde{A}^{(1)}-A^{(1)})d_1+B^{(0)}a_3-C^{(0)}a_2\,,\\
\label{general_BT_tt2}
a_{2t}&=&2A^{(0)}a_2+B^{(0)}(a_4-a_1)+d_2\widetilde{B}^{(1)}-d_1B^{(1)}\,,\\
\label{general_BT_tt3}
a_{3t}&=&-2A^{(0)}a_3+C^{(0)}(a_1-a_4)+d_1\widetilde{C}^{(1)}-d_2C^{(1)}\,,\\
\label{general_BT_tt4}
a_{4t}&=&-(\widetilde{A}^{(1)}-A^{(1)})a_4+C^{(0)}a_2-B^{(0)}a_3\,,
\ee
\end{itemize}
where $a_1,a_2,a_3,a_4$ are as in Proposition \ref{th}.
At this stage, it seems that there is an overdetermination since there are four equations for each part
whereas only two of each type are needed (the $x$ and $t$ transforms relating $\tq$ and $q$ and those relating
 $\tr$ and $r$). It turns out that half of them are indeed redundant.

\begin{proposition} {\bf B\"acklund transformations}\\
The B\"acklund transformations corresponding to $L$ are given 
by the equations for $a_2$ and $a_3$ in (\ref{general_BT_x2},\,\ref{general_BT_x3}) and (\ref{general_BT_t2},\,\ref{general_BT_t3})
or (\ref{general_BT_tt2},\,\ref{general_BT_tt3}).
The remaining equations for $a_1$ and $a_4$ can be deduced from them.
\end{proposition}
\prf
We know that the eigenvalues of $L^{(1)}$ are constant so
\be
(a_1a_4)_x=(a_2a_3)_x~~,~~a_{1x}+a_{4x}=0\,.
\ee
From this and the equations (\ref{general_BT_x2},\,\ref{general_BT_x3}) for $a_2$ and $a_3$, we deduce
\be
a_{1x}(a_4-a_1)=a_4(\tq a_3-r a_2)+a_1(\tr a_2-qa_3)\,.
\ee
Now, using $(\tq-q)a_3+(\tr -r)a_2=0$,
\be
(a_{1x}-\tq a_3+r a_2)(a_4-a_1)=0\,.
\ee
The possibility $a_4=a_1$ must be rejected in general since together with $a_{1x}+a_{4x}=0$ it would imply
that $a_1$ and $a_4$ are independent of $x$. Thus, we obtain the equation for $a_1$ and hence for $a_4$.

The proof for the $t$ part is similar. Useful identities in getting the result are obtained from
\be
\text{Tr}\,L_t=\text{Tr}\left[L(\tV-V)\right]=0\,,
\ee
and expanding in powers of $\lda$ or $\lda^{-1}$.
\finprf

We finish this general discussion by making a connection with Darboux matrices (see e.g. \cite{RS}).
Suppose that $\alpha_1\neq \alpha_2$ then we can define
\be
\label{def_P}
P=\frac{1}{\alpha_2-\alpha_1}(L^{(1)}-\alpha_1\1_2)\,,
\ee
and multiply $L(\lda)$ by $\frac{\lda}{\lda+\alpha_1}$ (since it is defined up to a function of $\lda$) to get
\be
\label{L_projector}
L(\lda)=\1_2+\frac{\alpha_2-\alpha_1}{\lda+\alpha_1}P\,.
\ee
The important fact is that $P$ is a projector ($P^2=P$ can be checked directly from the definition (\ref{def_P})). 
The form (\ref{L_projector}) for $L$ is usually encountered where the so-called B\"acklund-Darboux transformations
are used to generate multi-soliton solutions from a given one.
This form is also useful to exhibit the inverse of the B\"acklund matrix
\be
L^{-1}(\lda)=\1_2-\frac{\alpha_2-\alpha_1}{\lda+\alpha_2}P\,.
\ee

\section{Examples}\label{examples}

In this section, our results are applied on a variety of examples. We are able to reproduce the results of
\cite{BCZ,CZ} in a very simple way. For the NLS equation, we obtain a more general result
corresponding to a B\"acklund transformation with two real parameters (as it should, see e.g.
\cite{Chen}) instead of one. For the Korteweg-de Vries (KdV) and modified KdV equations, we obtain the 
defect contribution directly in terms of the original fields and reproduce the lagrangian approach
expressions in terms of "potential" fields. For each model, the defect conditions are given and consistently 
reproduces the associated well-known Backl\"und transformations, but taken at $x=x_0$ here, as expected
by construction.
We gather the examples in three classes according to certain symmetry considerations yielding information on
the defect parameters $\alpha_\pm$.

\subsection{Class I: $q=u$, $r=\epsilon u^*$, $\epsilon=\pm 1$, $u$ complex scalar field}

Let us introduce
\be
K=\left(
\begin{array}{cc}
0 & 1\\
\epsilon & 0
\end{array}\right)\,.
\ee
We have the following symmetries (we drop $x$ and $t$)
\be
U^*(\lda^*)=KU(\lda)K^{-1}~~,~~\tU^*(\lda^*)=K\tU(\lda)K^{-1}\,,
\ee
and we assume that $V$ and $\tV$ have the same properties. Therefore, we can look for the B\"acklund matrix
$L$ such that $L^*(\lda^*)=KL(\lda)K^{-1}$.
This implies $\alpha_2=\alpha_1^*$ so $\alpha_+\in\RR$ and $\alpha_-\in i\RR$. 
Then, it can be shown that the remaining conditions imply that the nontrivial
B\"acklund matrix reads
\be
\label{BM_classI}
L(\lda)=\1_2+\lda^{-1}\left(\begin{array}{cc}
\frac{1}{2}\left\{\alpha_+ \pm i\sqrt{\beta^2+\epsilon |\tu-u|^2} \right\} & -\frac{i}{2}(\tu-u)\\
\frac{i}{2}\epsilon(\tu^*-u^*) & \frac{1}{2}\left\{\alpha_+ \mp i\sqrt{\beta^2+\epsilon |\tu-u|^2} \right\}
\end{array}\right)\,,
\ee
where $\beta=i\alpha_-\in\RR$. For this class, the defect matrix, and hence the defect conditions
are parametrized by two arbitrary real numbers $\alpha_+$ and $\beta$.
In the case $\epsilon=-1$, we see that the transformation is such that $|\tu-u|^2\le \beta^2$.
We must take into account the fact that $r=\epsilon q^*$ in the discussion of the integrals of motion.
In particular, it turns out that to generate real integral of the motion (real classical
observables), one has 
to consider the following combination
\be
\label{Isym}
I^{sym}(\lda)=i(I(\lda)-I^*(\lda^*))\,.
\ee
So in practice, we will compute the contribution of the defect as
\be
I^{sym}_{defect}(\lda)=-i\left(\ln(L_{11}+L_{12}\Gamma)-\ln(L_{11}+L_{12}\Gamma)^*\right)|_{x=x_0}\,.
\ee
Now we expand (\ref{Isym}) in powers of $\lda^{-1}$ up to order 3 to illustrate the method. 
For convenience, we define $\Omega_\epsilon=\sqrt{\beta^2+\epsilon |\tu-u|^2}$.
At order $\lda^{-1}$, we find that the modified conserved density reads
\be
\label{order1}
\int_{-\infty}^{x_0}|\tu|^2 dx +\int^{\infty}_{x_0}|u|^2 dx \mp\epsilon\,\Omega_\epsilon|_{x=x_0}\,,
\ee
where the last term is the explicit defect contribution. Similarly,
at order $\lda^{-2}$, the modified conserved momentum is
\be
\label{order2}
\int_{-\infty}^{x_0}i(\tu\tu^*_x-\tu^*\tu_x) dx +\int^{\infty}_{x_0}i(uu^*_x-u^*u_x) dx 
-i\left[(u^*\tu-u\tu^*)\mp2\epsilon\beta\Omega_\epsilon\right]|_{x=x_0}\,.
\ee
Finally, the modified conserved energy is found to be
\be
\label{order3}
&&\int_{-\infty}^{x_0}(|\tu_x|^2+\epsilon|\tu|^4) dx +\int^{\infty}_{x_0}(|u_x|^2+\epsilon|u|^4) dx \nonumber\\
&&+\left[\mp\Omega_\epsilon(|\tu|^2+|u|^2)\mp\frac{\epsilon}{3}\Omega_\epsilon(3\beta^2-\Omega_\epsilon^2)-i\beta
(\tu u^*-u\tu^*)\right]|_{x=x_0}\,.
\ee
These results hold for any member of class I so in particular they hold for the cubic focusing ($\epsilon=-1$) 
or defocusing ($\epsilon=1$) nonlinear Schr\"odinger equation for
the complex scalar field $u$
\be
iu_t+u_{xx}=\epsilon|u|^2u\,,
\ee
and similarly for $\tu$. Indeed, this equation is obtained in the AKNS scheme by taking
\be
A(\lda)=-2i\lda^2+i|u|^2~~,~~B(\lda)=\epsilon C^*(\lda^*)=2\lda+iu_x\,,
\ee
and similarly for $\tu$.

Now the corresponding defect conditions can be derived from the general B\"acklund transformations given in the previous section. 
From the symmetry $a_3=\epsilon a_2^*$, we need only consider (\ref{general_BT_x2}) and (\ref{general_BT_t2}). 
Note that the $x$ part of the defect conditions is the same for all the models in class I. Here, for NLS,
we have at $x=x_0$
\be
\label{BTx_classI}
(\tu-u)_x&=&i\alpha_+(\tu-u) \pm (\tu+u)\sqrt{\beta^2+\epsilon |\tu-u|^2}\,,\\
(\tu-u)_t&=&-\alpha_+(\tu-u)_x\pm i(\tu+u)_x\sqrt{\beta^2+\epsilon |\tu-u|^2}+i(\tu-u)(|u|^2+|\tu|^2)\,.
\ee
Setting $\epsilon=-1$ and choosing the $-$ sign in (\ref{order1}), (\ref{order2}), (\ref{order3}), 
we note that we have to impose further $\beta=0$ to recover 
the results of \cite{CZ} (where the notation
$\Omega=\sqrt{\alpha^2-|\tu-u|^2}$ is used). This is due to the fact that the lagrangian
the authors took for the defect (the $\cB$ functional in their notations) is not the most general one. 
It corresponds to particular B\"acklund transformations with $\beta=0$.

\subsection{Class II: $q=u$, $r=\epsilon u$, , $\epsilon=\pm 1$, $u$ real scalar field}

This class is a subclass of class I with $u^*=u$ (and $\tu^*=\tu$). this immediately implies 
\be
\alpha_+=0\,.
\ee
The defect matrix for this 
class reads
\be
\label{BM_classII}
L(\lda)=\1_2+\lda^{-1}\left(\begin{array}{cc}
\pm\frac{i}{2}\sqrt{\alpha^2+\epsilon (\tu-u)^2} & -\frac{i}{2}(\tu-u)\\
\frac{i}{2}\epsilon(\tu-u) & \mp\frac{i}{2}\sqrt{\alpha^2+\epsilon (\tu-u)^2}
\end{array}\right)\,.
\ee
So we can simply use the result of the previous class, setting $u^*=u$ and $\tu^*=\tu$ (and thus, without
symetrizing). We exhibit the first orders for a few examples, taking advantage of specific forms of the defect matrix
in each case.

\subsubsection{Modified Korteweg-de Vries equation}

The modified Korteweg-de Vries equation equation 
\be
u_t-6\epsilon u^2u_x+u_{xxx}=0\,,
\ee
is obtained in the AKNS scheme by taking
\be
A(\lda)=-4i\lda^3-2i\lda\epsilon u^2~~,~~B(\lda)=\epsilon C^*(\lda^*)=4\lda^2 u+2i\lda u_x-u_{xx}+2\epsilon u^3\,,
\ee
and similarly for $\tu$. The modified conserved density reads
\be
\int_{-\infty}^{x_0}\tu^2 dx +\int^{\infty}_{x_0}u^2 dx\mp\epsilon\sqrt{\alpha^2+\epsilon (\tu-u)^2}|_{x=x_0}\,.
\ee
The modified conserved momentum is
\be
\label{integ_parts1}
\int_{-\infty}^{x_0}\tu \tu_x dx +\int^{\infty}_{x_0}uu_x dx -\frac{1}{2}(\tu^2-u^2+\alpha^2)|_{x=x_0}\,,
\ee
as can be directly checked by integration by parts (the constant $\alpha^2$ being irrelevant).
Finally, the next order yields
\be
\int_{-\infty}^{x_0}(\tu_x^2+\epsilon\tu^4) dx +\int^{\infty}_{x_0}(u_x^2+\epsilon u^4) dx
\mp\Omega_\epsilon\left[(\tu^2+u^2)-\frac{\epsilon}{3}\Omega_\epsilon^2\right]|_{x=x_0}\,,
\ee
where here $\Omega_\epsilon=\sqrt{\alpha^2+\epsilon (\tu-u)^2}$. The corresponding defect conditions read
\be
(\tu-u)_x&=&\pm (\tu+u)\sqrt{\alpha^2+\epsilon (\tu-u)^2}\,,\\
(\tu-u)_t&=&\pm\left\{2\epsilon(\tu^3+u^3)-(\tu+u)_{xx} \right\} \sqrt{\alpha^2+\epsilon (\tu-u)^2}\,.
\ee
It is worth noting that everything is expressed directly in terms of the fields $u$ and $\tu$. This
should be compared with the Lagrangian approach of \cite{CZ} where this was not possible. The use of "potential"
fields $p$ and $q$ such that $\tu=p_x$ and $u=-q_x$ is required in this formulation. 
Under this substitution, an alternative form
of the defect matrix can be derived
\be
L(\lda)=\1_2\pm\lda^{-1}\frac{i\alpha}{2}\left(\begin{array}{cc}
\cos (\tv-v)& -\sin (\tv-v)\\
-\sin(\tv-v) & -\cos(\tv-v)
\end{array}\right)\,.
\ee
and we consistently recover their result for the defect contribution to the first conserved quantity
 (setting $\epsilon=-1$), that is $\epsilon\sqrt{\alpha^2+\epsilon (\tu-u)^2}|_{x=x_0}$ becomes 
\be
\alpha(\cos(p-q)-1)|_{x=x_0}\,,
\ee
(a constant can always be added).

\subsubsection{Sine/sinh-Gordon equation}

This example illustrates the case $V$, $\tV$ polynomials in $\lda^{-1}$.
The sine-Gordon equation
in light-cone coordinates
\be
v_{xt}=\sin v\,,
\ee
is obtained by setting $u=-\frac{v_x}{2}$, $\epsilon=-1$ and taking
\be
A(\lda)=\frac{i\cos v}{4\lda}~~,~~B(\lda)=\frac{i\sin v}{4\lda}\,,
\ee
and similarly for $\tv$. For this model, the defect matrix takes the nice following form
\be
L(\lda)=\1_2\pm\frac{i\alpha}{2\lda}\left(\begin{array}{cc}
\cos\frac{\tv+v}{2} & -\sin\frac{\tv+v}{2}\\
-\sin\frac{\tv+v}{2} & -\cos\frac{\tv+v}{2}
\end{array}\right)\,,
\ee
where $\alpha$ is a nonzero real parameter.
Therefore, the modified conserved momentum reads 
\be
\frac{1}{4}\int_{-\infty}^{x_0}\tv_x^2 dx +\frac{1}{4}\int^{\infty}_{x_0}v_x^2 dx 
\pm \alpha\cos\frac{\tv+v}{2}|_{x=x_0}\,.
\ee
This is just the result in \cite{BCZ2} for the momentum and energy but expressed here in 
light-cone coordinates.
At the next order, as can be anticipated from (\ref{integ_parts1}), the bulk
contribution can be integrated by parts and combines nicely with the defect contribution, leaving 
the constant $-\frac{\alpha^2}{2}$
as the conserved quantity. Finally, the third order conserved quantity is
\be
\frac{1}{2}\int_{-\infty}^{x_0}(-\tv_{xx}^2+\frac{\tv_x^4}{4}) dx +\frac{1}{2}\int^{\infty}_{x_0}
(-v_{xx}^2+\frac{v_x^4}{4}) dx \pm\frac{\alpha}{3}\cos\frac{\tv+v}{2}(\tv_x^2+\tv_xv_x+v_x^2+2\alpha^2)|_{x=x_0}\,.
\ee
The defect conditions at $x=x_0$ are given by
\be
(\tv-v)_{x}&=& \pm 2\alpha\sin\frac{\tv+v}{2}\,,\\
(\tv+v)_{t}&=& \pm\frac{2}{\alpha}\sin\frac{\tv-v}{2}\,.
\ee

\subsubsection{Liouville equation}

This is another example of evolution equation obtained from $V$ and $\tV$ polynomials in 
$\lda^{-1}$.
We need to discuss this equation in detail since, as mentioned above, the condition
of vanishing fields at infinity and the boundary conditions (\ref{BC}) are not applicable. However, 
it is straightforward to prove that Proposition \ref{th} is still valid. Also, 
it is still possible to prove Proposition \ref{prop_conserved} with a slight modification, as we now show. 
The Liouville equation 
in light-cone coordinates for the field $v$
\be
v_{xt}=2e^{v}\,,
\ee
is obtained by taking
\be
u=\frac{v_x}{2}~~,~~\epsilon=1~~,~~
A(\lda)=\frac{i e^v}{2\lda}=-B(\lda)\,,
\ee
and similarly for $\tv$. The conservation laws (\ref{conservation1}) and (\ref{conservation2}) still hold since
they do not depend on the boundary conditions. But now, (\ref{intermed}) becomes
\be
\partial_t\int_{x_0}^\infty q\Gamma dx+\partial_t\int^{x_0}_{-\infty}
\tq\widetilde{\Gamma}dx&=&\lim_{x\to\infty}(B\Gamma+A)-\lim_{x\to-\infty}(\widetilde{B}
\widetilde{\Gamma}+\widetilde{A})\\
&&+\left(\widetilde{B}
\widetilde{\Gamma}+\widetilde{A}-(B\Gamma+A)\right)|_{x=x_0}\,.
\ee
The last term in the right-hand-side is treated as before. The point is to recast the other two terms as
time derivatives. For this equation, one has $A=C=-B$ so $B\Gamma+A=A(1-\Gamma)$ and the Ricatti equation (\ref{ricatti_t})
becomes
\be
\Gamma_t=A(1-\Gamma)^2\,.
\ee
Therefore, since $\Gamma\neq 1$ (this can be seen from (\ref{Ricatti_x}))
\be
B\Gamma+A=-\partial_t\ln(1-\Gamma)\,.
\ee
The same result holds for $\widetilde{\Gamma}$. Therefore, Proposition \ref{prop_conserved} is modified
to the following.

The generating function for the integral of motions of the Liouville equation reads
\begin{eqnarray}
I(\lda)&=&I_{bulk}^{left}(\lda)+I_{bulk}^{right}(\lda)+I_{defect}(\lda)\,,
\ee
where\be
I_{bulk}^{left}(\lda)&=&\frac{1}{2}\int^{x_0}_{-\infty}
\tv_x\widetilde{\Gamma}dx-\lim_{x\to-\infty}\ln(1-\widetilde{\Gamma})\,,\\
I_{bulk}^{right}(\lda)&=&\frac{1}{2}\int_{x_0}^\infty v_x\Gamma dx+\lim_{x\to\infty}\ln(1-\Gamma)\,,\\
I_{defect}(\lda)&=&-\ln (L_{11}+L_{12}\Gamma)|_{x=x_0}\,,
\end{eqnarray}
and $L_{ij}$'s are the entries of the B\"acklund matrix $L$. The additional contributions essentially
kill terms arising from trivial integration by parts in the integrals of motion. 

The defect matrix can be written
\be
L(\lda)=\1_2+\frac{i\gamma}{4\lda}\, e^{\pm\frac{\tv+v}{2}}\left(\begin{array}{cc}
 1 & -1\\
1 & -1
\end{array}\right)\,,
\ee
where $\gamma$ is a nonzero real constant. At first order, the modified conserved momentum reads
\be
\frac{1}{2}\int^{x_0}_{-\infty}
\tv_x^2dx+\tv_x|_{-\infty}+\frac{1}{2}\int_{x_0}^\infty
v_x^2dx-v_x|_{\infty}-\gamma e^{\pm\frac{\tv+v}{2}}|_{x=x0}\,.
\ee
As is now customary, the next order combines nicely to produce
\be
(v_x^2-2v_{xx})|_{\infty}-(\tv_x^2-2\tv_{xx})|_{-\infty}\,,
\ee
which can be checked directly to be a constant. Finally, at the third order, we have
\be
&&\frac{1}{2}\int^{x_0}_{-\infty}
\left(\tv_{xx}^2+\frac{\tv_x^4}{4}\right)dx-\left(\tv_{xxx}-\frac{\tv_x^3}{6}-\tv_x\tv_{xx} \right)|_{-\infty}\nonumber\\
&+&\frac{1}{2}\int_{x_0}^\infty
\left(v_{xx}^2+\frac{v_x^4}{4}\right)dx+\left(v_{xxx}-\frac{v_x^3}{6}-v_xv_{xx} \right)|_{\infty}\nonumber\\
&-&\frac{\gamma}{6}\,e^{\pm\frac{\tv+v}{2}}(\tv_x^2+v_x^2+\tv_xv_x)|_{x=x0}\,.
\ee
The defect conditions are given by
\be
(\tv-v)_{x}&=&\gamma\, e^{\pm\frac{\tv+v}{2}}\,,\\
(\tv+v)_{t}&=&\pm \frac{4}{\gamma}\,e^{\mp\frac{\tv+v}{2}}(e^\tv-e^v)\,,
\ee

\subsection{Class III: $q=u$, $r= \epsilon$, , $\epsilon=\pm 1$, $u$ real scalar field}

For this class, there is no special symmetry. The derivation of the defect matrix deserved special attention for this class.
Indeed, if we assume that $(\tU,\tV)$ has the same form
as $(U,V)$, as we have done so far, then an immediate consequence is $a_3=0$ and
$a_1=a_4=\alpha_1$ are constant. From this, equation (\ref{general_BT_x1}) implies $\tu=u$ \ie we get
the trivial defect conditions. A solution to this problem is to left multiply $L$ by $\sigma_3$ or
equivalently to consider $(\sigma_3\tU\sigma_3,\sigma_3\tV\sigma_3)$. This amounts to change the sign of the 
off-diagonal terms. Then, equation (\ref{general_BT_x3}) implies $\alpha_+=0$. Finally, taking into account the reality
of the field, one gets the defect matrix for this class
\be
\label{BM_classIII}
L(\lda)=\1_2+\lda^{-1}\left(\begin{array}{cc}
\pm\frac{i}{2}\sqrt{\beta^2+2\epsilon (\tu+u)} & \frac{i}{2}(\tu+u)\\
-i\epsilon & \mp\frac{i}{2}\sqrt{\beta^2+2\epsilon (\tu+u)}
\end{array}\right)\,,
\ee
We apply our method to the Korteweg-de Vries equation
\be
u_t-6\epsilon uu_x+u_{xxx}=0\,,
\ee
which is obtained in the AKNS scheme by taking
\be
A(\lda)&=&-4i\lda^3-2i\epsilon\lda u+\epsilon u_x\,,\\
B(\lda)&=&4\lda^2 u+2i\lda u_x+2\epsilon u^2-u_{xx}\,,\\
C(\lda)&=&4\epsilon \lda^2+8 u\,,
\ee
and similarly for $\tu$, with appropriate change of signs. We make direct use of (\ref{generating}). 
The first nontrivial order is 
$\lda^{-3}$ and for the modified conserved density reads
\be
\frac{1}{2}\int_{-\infty}^{x_0}\tu^2 dx +\frac{1}{2}\int^{\infty}_{x_0}u^2 dx
\mp \frac{1}{6}\sqrt{\beta^2+2\epsilon (\tu+u)}\left(\epsilon(\tu+u)-\beta^2  \right)|_{x=x_0}\,.
\ee
The defect conditions are given by
\be
(\tu+u)_x&=&\pm(\tu-u)\sqrt{\beta^2+2\epsilon (\tu+u)}\,,\\
(\tu+u)_t&=&\pm\left(3\epsilon(\tu^2-u^2)-(\tu-u)_{xx}\right)\sqrt{\beta^2+2\epsilon (\tu+u)}\,.   
\ee

Once again, we note that everything is expressed directly in terms of the initial fields $u$ and
$\tu$ while this was not possible in the lagrangian approach. It is easy to make contact with the latter
by setting $u=q_x$ and $\tu=p_x$. Then, 
\be
(p+q)_x=\frac{\epsilon}{2}\left[(p-q)^2-\beta^2\right]\,.
\ee
Taking $\epsilon=1$ and setting $\beta^2=-4\alpha$, we recover the result of \cite{CZ} for the defect contribution to the
density (the momentum in their setting)
\be
\frac{1}{2}\int_{-\infty}^{x_0}p_x^2 dx +\frac{1}{2}\int^{\infty}_{x_0}q_x^2 dx \,,
\ee
that is
\be
\left(-\alpha(p-q)-\frac{1}{12}(p-q)^3\right)|_{x=x_0}\,.
\ee

\subsection{Remarks}
We will not go into the analysis of the higher $N$ case in (\ref{truncate}). We simply note that a large class
of higher $N$ defect matrices is provided by products of $N=1$ defect matrices. The corresponding
defect conditions are then simply compositions of the defect conditions we derived above.
In other words, the defect matrices we constructed have a group structure, as can be seen directly from 
(\ref{eq_diff_L}) and (\ref{eq_diff_L2}).
Furthermore, Bianchi's theorem of permutability (see e.g. \cite{RS}) implies that this is an abelian group.
An important question concerns the existence and properties of those higher $N$ defect matrices which do not factorize
as products of $N=1$ ones. Their study would shed new light on possible new defect conditions for the well-known systems
we discussed. Also, the question of defect matrices preserving integrability but which do not fall at all in the class 
discussed here remains entirely open. In this sense, no claim of uniqueness of defect matrices is made and one should 
remember that the proposed approach here is sufficient to ensure integrability in the presence of a defect. The issue
of finding necessary defect conditions for integrability is not answered.
It should also be noted that following the linearization argument of \cite{BCZ,CZ}, it appears that the defect
conditions constructed here allow for pure transmission only. This has been checked explicitely for all the examples
given above.

\section{Extension to another scheme}

The Kaup-Newell (KN) \cite{KN} scheme goes along the same steps as the AKNS scheme to produce integrable evolution
equations with the essential difference that the matrix $U$ involved in the $x$ part of the auxiliary problem
(\ref{pb_aux}) has the following form
\be
U_{KN}=\left(\begin{array}{cc}
-i\lda^2 & \lda q\\
\lda r & i\lda^2
\end{array}\right)=-i\lda^2\sigma_3+\lda W\,.
\ee
One well-known model obtained in this scheme is the derivative nonlinear Schr\"odinger equation
\be
iu_t+u_{xx}=2\epsilon(|u|^2u)x\,,
\ee
where $q=u=\epsilon r$, $\epsilon=\pm 1$ and the matrix $V_{KN}$ should read
\be
V_{KN}=\left(\begin{array}{cc}
-2i \lda^4-i\epsilon\lda^2 |u|^2 & 2\lda^3 u+i\lda u_x+\epsilon\lda |u|^2u\\
2\epsilon\lda^3 u^*-i\epsilon\lda u^*_x+\lda |u|^2u^* & 2i \lda^4+i\epsilon\lda^2 |u|^2
\end{array}\right)\,.
\ee
It turns out that our method works for this scheme too with the appropriate modifications.
We proceed as before by taking two copies of the auxiliary problem related by a matrix 
$M$ such that
\begin{eqnarray}
\label{eq_diff_M}
M_x&=&\widetilde{U}_{KN}M-MU_{KN}\,,\\
\label{eq_diff_M2}
M_t&=&\widetilde{V}_{KN}M-MV_{KN}\,.
\end{eqnarray}
Then we assume that the two copies are related by $M$ at some point $x=x_0$.
\begin{proposition}
The generating function for the integral of motions reads
\begin{eqnarray}
I(\lda)&=&I_{bulk}^{left}(\lda)+I_{bulk}^{right}(\lda)+I_{defect}(\lda)\,,
\ee
where\be
I_{bulk}^{left}(\lda)&=&\int^{x_0}_\infty
\lda\tq\widetilde{\Gamma}dx\,,\\
I_{bulk}^{right}(\lda)&=&\int_{x_0}^\infty \lda q\Gamma dx\,,\\
I_{defect}(\lda)&=&-\ln (M_{11}+M_{12}\Gamma)|_{x=x_0}\,,
\end{eqnarray}
and $M_{ij}$'s are the entries of the B\"acklund matrix $M$. For this scheme, 
$\Gamma$ and $\widetilde{\Gamma}$ have a different expansion as $\lda\to\infty$
\be
\Gamma=\sum_{n=0}^\infty\frac{\Gamma_n}{(2i\lda)^{2n+1}}\,,
\ee
with
\be
\Gamma_0=-r~~,~~\Gamma_{n+1}=2i\Gamma_{nx}+q\sum_{p=0}^n\Gamma_p\Gamma_{n-p}\,,
\ee
and similarly for $\widetilde{\Gamma}$.
\end{proposition}
\prf
The proof is the same as that of Proposition \ref{prop_conserved}, the only difference being that
the conservation equations now read
\begin{eqnarray}
\left(q\Gamma\right)_t&=&\frac{1}{\lda}\left(B\Gamma+A\right)_x~~,~~\forall x>x_0\,,\\
\left(\tq\widetilde{\Gamma}\right)_t&=&\frac{1}{\lda}\left(\widetilde{B}
\widetilde{\Gamma}+\widetilde{A}\right)_x~~,~~\forall x<x_0\,,
\end{eqnarray}
due to the different $\lda$ dependence of $U_{KN}$. The latter is also responsible for
the different series expansions of $\Gamma$ and $\widetilde{\Gamma}$ coming from the following Ricatti
equation
\be
\Gamma_x=\lda r+2i\lda^2\Gamma-\lda q\Gamma^2\,,
\ee
and similarly for $\widetilde{\Gamma}$.
\finprf

To apply this to specific models, we would need some sort of classification of the matrices $M$ along the lines of what
is available for $L$. However, the author is not aware of such results. It is a problem for future investigation.

\section{Discussion of integrability}\label{discussion}

So far, we have shown how the infinite set of conservation laws is modified by the presence of 
a defect described by a matrix for any evolution equation falling into the AKNS or KN schemes.
The role of an infinite set of conserved quantities is well-known in the construction of action-angle variables
in the inverse scattering method for evolution equations on the line. In turn, this construction ensure the integrability
of the system in the sense that in terms of the new variables, the evolution in time is very easy to solve. Then, using
the inverse part of the method (the Gelfan'd-Levitan-Marchenko equations \cite{GL,Mar}) one can deduce the time evolution of the original
fields (see e.g. \cite{AKNS}). So, in this sense, the systems with defect we have considered are integrable\footnote{For completeness,
one should perform the inverse scattering method and identify the action-angle variables in this context. This is left
for future work.}.

A traditional complementary view is to reformulate the evolution equation as Hamiltonian systems with a 
Poisson structure. In this formalism, the idea is to show that the conserved quantities previously constructed
form a commutative Poisson algebra containing the Hamiltonian which generates the time evolution of 
the fields. Then, one talks about integrability in the sense of Liouville. One of the advantages of this reformulation
is the possibility of quantization and this was the object of the quantum inverse scattering method, see e.g. \cite{FT}.

The method of the classical $r$-matrix \cite{Skly_r} has proved very useful and fundamental in
the discussion of these issues for classical integrable systems on the whole line. 
The situation on the half-line is also well understood \cite{Skly_b}. Our aim in this section is to
discuss the situation with a defect. 

Let us first recall some facts about the method of the classical $r$-matrix for ultralocal models. The basic ingredient is the
$2\times 2$ transition matrix $T(x,y,\lda)$, $x<y$, defined as the fundamental solution
 of the $x$-part of the auxiliary problem
at a given time (not explicitely displayed in T)
\be
\partial_y T(x,y,\lda)=U(y,t,\lda)T(x,y,\lda)~~,~~T(x,x,\lda)=\1_2\,,
\ee
where $U$ is given as in (\ref{lax_pair}).
The important result reads (see e.g. \cite{Fad})
\be
\label{finite_rtt}
\{T_1(x,y,\lda),T_2(x,y,\mu)\}=[r_{12}(\lda-\mu),T_1(x,y,\lda)~T_2(x,y,\mu)]\,,
\ee
where $r_{12}(\lda)$, the classical $r$-matrix, is a $4\times 4$ antisymmetric solution of the classical
Yang-Baxter equation \cite{BD}. For systems on the circle, it allows to show that the coefficients of the series expansion of $TrT(x,y,\lda)$ in
$\lda$ are in involution and the Hamiltonian is one of the them.
For systems on the line, the principle is the same but one has to take the infinite
volume limit of (\ref{finite_rtt}), usually with special care.

For the problem with defect, we would like to mimic this procedure. The main difference here is that there is something
nontrivial going on at a single point $x=x_0$ and characterized by $L$ (or $M$\footnote{For convenience, in the rest of the paper, 
we stick to a single notation $L$ for the defect matrix.}). The standard procedure becomes ill-defined in the continuous case as it involves
singular Poisson brackets of the type "$\delta(0)$" where $\delta(x)$ is the Dirac distribution.

Indeed, the analog of the transition matrix for $x<x_0<y$ is 
\be
\label{def_mono_defect}
\cT^{x_0}(x,y,\lda)=T(x_0,y,\lda)L^{-1}(x_0,t,\lda)\widetilde{T}(x,x_0,\lda)\,,
\ee
and the task of computing 
\be
\{\cT^{x_0}_1(x,y,\lda),\cT^{x_0}_2(x,y,\mu)\}\,,
\ee
involves computing $\{L^{-1}_1(x_0,t,\lda),L_2^{-1}(x_0,t,\lda)\}$.
There does not seem to be a direct approach starting from the explicit
form of $L$ as classified in this paper. However, in the context of finite-dimensional integrable
systems, important results were obtained by E. Sklyanin in \cite{Skly_cano} concerning the canonicity of 
B\"acklund transformations in the formalism of the $r$-matrix approach. Transposing the results in the present context
and assuming we have the same theories on both sides of the defect, \ie $T(x,y,\lda)$ and 
$\widetilde{T}(x',y',\lda)$ satisfy (\ref{finite_rtt}), for $x,y>x_0$ and $x',y'<x_0$ respectively, it seems reasonable to 
postulate that $L^{-1}$ is just 
another representation of the Poisson algebra (\ref{finite_rtt}) so that
\be
\label{L_rtt}
\{L^{-1}_1(x_0,t,\lda),L_2^{-1}(x_0,t,\lda)\}=[r_{12}(\lda-\mu),L^{-1}_1(x_0,t,\lda)~L_2^{-1}(x_0,t,\lda)]\,.
\ee
From this it immediately follows that
\be
\{\cT^{x_0}_1(x,y,\lda),\cT^{x_0}_2(x,y,\mu)\}=
[r_{12}(\lda,\mu),\cT^{x_0}_1(x,y,\lda)~\cT^{x_0}_2(x,y,\mu)]\,.
\ee
Note that the 
generalization to $N$ defects can be described in this formalism as well.
Using the same notations as Section \ref{several}, one construct the monodromy matrix
\be
\cT^N(\lda)&=&T^{N+1}(x_N,x_{N+1},\lda)(L^N)^{-1}(x_N,t,\lda)\dots
(L^2)^{-1}(x_1,t,\lda)T^1(x_1,x_2,\lda)\nonumber\\
&&\qquad\qquad\qquad\qquad\qquad\times(L^1)^{-1}(x_1,t,\lda)T^0(x_0,x_1,\lda)\,,
\ee
with all the $L^j$'s satisfying (\ref{L_rtt}).

This discussion brings us to the connection between the lagrangian approach of \cite{BCZ,CZ} to integrable defects and a
quite standard procedure to implement inhomogeneities or impurities in discrete integrable systems. 
Indeed, through the approach of this paper, 
one can
reformulate the lagrangian approach in terms of a transition matrix made of two bulk parts and a localised defect part
realising a different representation of the same Poisson algebra. But this is exactly what is usually done to 
implement so-called impurities or inhomogeneities in discrete integrable systems. 
Let us consider a discrete ultralocal system of length
$\ell$ with $N$ sites.
The transition matrix $T(\lda)$
is then a product of local matrices $t_j(\lda)$, $j=1,\dots,N$ satisfying
\be
\label{classical_rtt}
\{t_{j0}(\lda),t_{k0'}\}=\delta_{jk}\left[r_{00'}(\lda-\mu),t_{j0}(\lda)t_{k0'}(\mu) \right]\,,
\ee
in the same representation (the index $j$ represents the space of dynamical variables while $0$, $0'$ are
auxiliary spaces, $\CC^2$ here). One can check then that $T(\lda)=t_N(\lda)\dots t_1(\lda)$ satisfies
\be
\label{classical_rtt2}
\{T_{0}(\lda),T_{0'}(\mu)\}=\left[r_{00'}(\lda-\mu),T_{0}(\lda)T_{0'}(\mu) \right]\,,
\ee
\ie exactly (\ref{finite_rtt}).
To introduce an inhomogeneity at site $j_0$ say, one then
chooses a different representation $\hat{t}_{j_0}(\lda)$ of the same algebra. This does not change the properties
of $T(\lda)=t_N(\lda)\dots \hat{t}_{j_0}(\lda)\dots t_1(\lda)$ but influences the physical quantities (e.g. the
integrals of motion) that can be computed since
the latter depend on the representation at each site. Again, the introduction of several inhomogeneities
can be done straightforwardly. This actually provides a solution to the above problem of computing Poisson brackets of
defect matrices by considering lattice regularizations
of integrable field theories and changing representations appropriately at local sites to generate defects. 

From this point of view, one can anticipate the outcome of the quantization of this approach.
It is known that (\ref{classical_rtt2}) is the classical ($\hbar\to 0$) limit of the quantum Yang-Baxter 
algebra
\be
R_{00'}(\lda-\mu)\tau_{j0}(\lda)\tau_{j0'}(\mu)=\tau_{j0'}(\mu)\tau_{j0}(\lda)R_{00'}(\lda-\mu)\,,
\ee
where $R(\lda)$ is the quantum $R$ matrix associated to $r$
\be
R(\lda)=\1+i\hbar r(\lda)+O(\hbar^2)\,,
\ee
and $\tau_j(\lda)\to t_j(\lda)$ in the $\hbar\to0$ limit.
So the quantum defect matrix $\cL(\lda)$ encoding
the defect conditions will satisfy the quantum Yang-Baxter algebra. This gives some support to the ad hoc
quantization procedure adopted in \cite{BCZ2} for the sine-Gordon with integrable defect. 
We would like to stress that the above programme of discretization has been completed for the defect sine-Gordon model in the important paper
\cite{KH}, both at classical and quantum level. The approach is based on the notion of ancestor algebra \cite{Kun}.

\section*{Conclusions and outlook}

In this paper, we have reformulated the lagrangian approach to the question of integrable defects in the 
language of the inverse scattering method, taking advantage of the common features that had been observed on
a case by case study: frozen B\"acklund transformations as defect conditions ensure integrability. The reformulation allows a 
systematic proof of this as well as an efficient computation of the modified conserved quantities to all orders in 
terms of the defect matrix. The latter, and the associated defect conditions, can be classified 
and we performed these computations for a certain class of matrices. Taking particular examples, we recovered and
even generalized all the previous results obtained by the lagrangian method. It should be emphasized that 
this procedure provides a sufficient approach to the question of integrable defects in classical field theories and,
by no means, represents a complete picture of the story.

Rather, it is a first step for future developments among which further study of the classical $r$ matrix approach and 
quantization of the method are important. Let us mention also the contruction of other integrable defects allowing 
if possible reflection as well. If applicable, the quantization should
then be related to existing quantum algebraic frameworks like the Reflection-Transmission algebras \cite{MRS}. Finally,
the complete setup of the direct and inverse part of the method for the actual construction of the solutions, 
especially of soliton type, should shed new light
on the results already obtained by the more direct approach of \cite{BCZ,CZ}.

\section*{Acknowledgements}
It is a pleasure to thank E. Sklyanin and E. Corrigan
for discussions and encouragements in the course of this paper. We also warmly thank E. Ragoucy for
useful comments in the final stage of this work.


\begin{thebibliography}{99}
\bibitem{NB} The initial data is typically chosen to be infinitely smooth in the $x$ variable in the bulk but the defect conditions ensure only
that derivatives of the fields up to a finite degree are well-behaved as $x\to x_0^\pm$. 
For each higher degree of "smoothness", one has
to impose additional conditions at the point of the defect which can be obtained by differentiating for instance (\ref{general_BT_x1}-\ref{general_BT_x4}), 
using all known relations
to get an expression involving only the fields and their derivatives up to the desired degree, and finally set $x=x_0$. 
This is important to make sense of the higher conserved quantities.
\end{thebibliography}
\end{document}